\documentstyle[aps,prb]{revtex}
\begin{document}

\title{Theory of polarized Fermi liquid} 
\author{V.P.Mineev}
\address{Commissariat a l'Energie Atomique, DSM, Departement de
Recherche Fondamentale sur la Matiere Condensee, SPSMS, 38054
Grenoble, France } 
\date{9 october 2004}
\maketitle
\begin{abstract}

The dispersion law of transverse spin waves known in the Stoner-Hubbard
model of itinerant ferromagnetism corresponds to that is well known in
more broder and well controlled approach of Fermi-liquid theory. 
Making use the quantum-field theoretical approach we derive the
dispersion law for the transverse spin waves in a weakly polarized
Fermi liquid at $T=0$.  Along with the dissipationless part inversely
proportional to the polarization it contains also the finite
zero-temperature damping.  It is shown that similar derivation for
"ferromagnetic Fermi liquid" taking into consideration the divergency
of static transverse susceptibility also leads to the same attenuating
spin wave spectrum.  Hence, in both cases we deal in fact with spin
polarized Fermi liquid but not with isotropic itinerant ferromagnet
where the zero temperature atenuation is prohibited by Goldstone
theorem.  It demonstrates, the troubles of the Fermi liquid
formulation of a theory of itinerant ferromagnetic systems.
\end{abstract}

\bigskip

\section{Introduction}

It is widely accepted that the Hubbard-Stoner model serves a sort of
baseground for the description of the itinerant ferromagnetism \cite{1}.  
The exchange interaction between the electrons is just ignored here but
instead the model deals with the contact or extremely short range
repulsion between the fermi particles with the opposite spins.  The latter 
leads to the ground state with finite polarization described by the
wave function
\begin{equation}
\Psi_{0}¥=\prod_{p<p_{+}¥}¥a^{\dagger}¥_{p\uparrow}¥
\prod_{p'<p_{-}¥}¥a^{\dagger}¥_{p'\downarrow}¥\Psi_{vac}¥,
\label{ea}
\end{equation}
where $p_{+}¥$ and $p_{-}¥$ are the Fermi surface radia with spin
directions parallel and antiparallel to the magnetization direction. 
The elementary excitations in the model are the excitations of paires
electron and hole with opposite spins.  There is also a collective
mode which is the transverse spin waves.  Usually it is thought that
the Stoner model gives not bad qualitative description of itinerant
ferromagnetism but for the creation of more realistic quantitative
theory it must be corrected by taking into account of spin-wave
-electron-hole interaction.  At the same time it is customary accepted
to shut our eyes to the qualitative consequences of Stoner model which
look strange for a ferromagnet.

First, it is the Stoner criterium of ferromagnetism
according to which the ferromagnetic state appears at high enough
interactions and density of states of Fermi gas $UN_{o}¥ >1$.  The latter in
the case of $D=3$ one band electron system with quadratic dispersion
law means the existance of ferromagnetism at high enough densities of
electron gas.  This property of Hubburd model with short range
interaction is just the opposite to the property of electron gas with
Coulomb exchange interaction (see for example the book \cite{2}) where it
is easy to check that the ferromagnetic state appears in the low
density limit as it has been pointed out by F.Bloch \cite{3}.  The
criterium of ferromagnetism looks here as
$e^{2}¥m>\hbar^{2}¥n^{1/3}¥$.

Secondly, the reactive part of diffusion constant $\Re D$ in quadratic
spin wave dispersion law
\begin{equation}
\omega=Dk^{2}¥
\label{eb}
\end{equation}
obtained in frame of Hubbard-Stoner model \cite{1} is proved
inversionally proportional to the magnetization
\begin{equation}
\Re D \propto \frac{1}{{\bf M}}\propto\frac{1}{p_{+}¥-p_{-}¥}.
\label{ec}
\end{equation}
Whereas according to Landau and Lifshits \cite{4} the magnetization in any
ferromagnet obeys the equation of motion
\begin{equation}
\frac{\partial {\bf M}}{\partial t}=K\nabla^{2}¥{\bf M}\times{\bf M}.
\label{ed}
\end{equation}
Hence the diffusion constant in the spin wave dispersion law is proportional to
magnetization that physically corresponds to the finite rigidity of
the ferromagnetic ordering preventing formation of inhomogenious
states.

Thirdly, unlike to Landau-Lifshits spin wave dispesion law the Stoner
dispersion always contains a dissipation discribed by the imaginary part of 
diffusion constant.  The presence of dissipation at $T=0$ in an ordered state
contradicts to the Goldstone theorem.

These properties having nothing common with ferromagnetism
demonstrate the difficulties of Stoner-Hubbard model which does not
take into account the many particle character of ferromagnetic ground state
and rather related to the polarized Fermi gas with wave function (\ref{ea}).
The same properties are well known  for spin polarized neutral Fermi
liquid like liquid $^{3}¥He$ or $^{3}¥He-^{4}¥He$ solutions.  They can
be obtained in frame of more broader and well controlled approach of
the theory of Fermi liquid with short range interaction between
particles \cite{5,6,7,8,9,10}.

On the other hand there were known also several publications with pretension to
develope a theory of genuine itinerant isotropic ferromagnet based on
Landau Fermi liquid theory \cite{11,12,13}. In particular the
derivation of dissipationless (up to the terms of $k^{4}¥$ power) spin
waves spectrum with diffusion constant proportional to magnetization
has been announced.

The goal of the present article is to reconsider in
frame of microscopic theory the problem of transverse spin waves in
spin-polarized Fermi liquid and in "itinerant ferromagnet" as it has
been defined in the papers \cite{11,12,13}.  It is shown that in the
both cases the microscopic derivation leads to the same spin wave
spectrum.  Along with the dissipationless part inversely proportional
to the polarization it contains also the finite zero-temperature
damping.  The polarization dependence both dissipative and reactive
part of diffusion constant corresponds to dependences found earlier by
means of kinetic equation with two-particle collision integral
\cite{10}.  It is shown that similar derivation for "ferromagnetic
Fermi liquid" taking into consideration the divergency of static
transverse susceptibility also leads to the same attenuating spin wave
spectrum.  Hence, in both cases we deal in fact with spin polarized
Fermi liquid but not with isotropic itinerant ferromagnet where the
zero temperature atenuation is prohibited by Goldstone theorem.  It
demonstrates, the troubles of the Fermi liquid formulation of a theory
of itinerant ferromagnetic systems.

In section II starting from the integral equation for the vortex
function for the scattering of two particles with opposite spin direction
we derive the Landau-Silin kinetic equation for the transversal spin waves
and find its dispersion law.
In the third section the latter is done by another way taking into
consideration the divergency of static transverse susceptibility.  The
paper is concluded by Conclusion Section.

\section{Spin waves in polarized Fermi liquid}

The problem of transverse spin-waves in spin-polarized Fermi liquid
has the long story.  The calculations of transverse spin-diffusion
coefficient in dilute degenerate Fermi gas with arbitrary polarization
have been done for the first time in the papers by W.Jeon and W.Mullin
\cite{5} where the low temperature saturation of corresponding
relaxation time has been established.  About the same time
A.Meyerovich and K.Musaelyan \cite{6,7} have generalized the Landau
derivation \cite{14} of Fermi liquid kinetic equation from microscopic
theory on the case transverse spin kinetics in the polarized Fermi
liquid and also come to the same conclusion.  A derivation and an
exact solution of the kinetic equation in the $s$-wave scattering
approximation for dilute degenerate Fermi gas with arbitrary
polarization at $T=0$ and for a small polarization at $T\ne 0$ have
been obtained also in the papers \cite{8} by D.Golosov and
A.Ruckenstein.  For the treatment of this problem in a Fermi liquid
the Matthiessen-type rule arguments and simple relaxation-time
approximation for the collision integral have been used \cite{9}. 
More recently the derivation of transverse spin dynamics in a
spin-polarized Fermi liquid from Landau-Silin kinetic equation with
general form of two-particle collision integral has been performed
\cite{10}.  The existance of zero-temperature damping of transverse
spin waves has been established.

Experimentally the saturation of the transverse spin wave
diffusion constant at temperatures about several millikelvin has been
registered by spin-echo technique (see for instance \cite{15}).  On the
other hand the spin wave experiments demonstrate the behaviour
characterized rather by the absence of transverse spin wave damping in
the same temperature region \cite{16}.  The latter seemed to be a
confirmation of the point of view of I.Fomin \cite{17} who has
argued for the dissipationless form of transverse spin wave spectrum
being obtained from the correction to the systems energy due to the
gauge transformation into the coordinate system where the
magnetization vector is a constant.  The same trick has been used
earlier for the treatment of one-particle and collective excitations
dualism in the itinerant ferromagnets by R.Prange and co-workers
\cite{18}, which is in our opinion still unresolved problem (see
below).  The calculation of the generalized susceptibility coefficient
in the expression for the spin current found in \cite{17} has not been
performed, just the reference on such the calculation \cite{19} in
superfluid $^{3}He$ had been given.  Indeed, one can calculate
susceptibility using the similar procedure.  However in the case of
polarized Fermi liquid one must use the Green functions with the
finite imaginery self energies parts due to collisions between
quasiparticles as it was done in \cite{7,8}, that inevitably leeds to
the spin waves attenuation.

I.Fomin also used an additional
argument in support of absence of attenuation of transversal
spin-waves in spin polarized Fermi liquid.  This was an analogy with
ferromagnetic Fermi liquid where were shown by P.Kondratenko 
\cite{12} that attenuation in the spin wave
spectrum arises only in the terms of order $\sim k^{4}¥$.  Indeed, 
it seems, that the space-time evolution of somehow artificially created 
magnetization in paramagnetic Fermi liquid in absence of spin-nonconserving
interactions
will be developed according the same laws as in isotropic itinerant
ferromagnet.  In reality this is not true.  The conservation of
magnetization does not accomplish the dissipationless dispersion of
magnons.  Even in the inhomogeneously rotating coordinate system,
where the magnetization vector is a constant, the quasiparticle
distribution function of paramagnetic Fermi liquid still is time and
coordinate dependent matrix in the spin space containing odd in
momentum off-diagonal part producing the spin current relaxation.  In
case of a ferromagnet the corresponding redistribution of the
particles over the states with different momenta and spin up and down
directions is prevented by the rigidity of many electron orbital wave
function.  The last property is not taken into account in the theory
of polarized Fermi liquid and all the attempts to discuss the
itinerant ferromagnets as sort of polarized Fermi liquid are
incomplete.

The Landau-type derivation of transverse spin
dynamics in a weakly spin-polarized Fermi-liquid from microscopic
theory has been performed in the paper \cite{7}.  Here we make a
similar derivation with the purpose to stress the conditions it needs
to be valid, to compare the answer with that had been obtained from
kinetic equation at nonzero temperatures \cite{10}, and to juxtapose this with
derivation for "ferromagnetic" Fermi-liquid \cite{13} which we also
reproduce after.

As in original paper by Landau \cite{14} we shall consider here a system of
fermions at $T=0$, with arbitrary short range interaction forces.  The
presence of polarization means that the particle distribution
functions for spin-up and spin-down particles have different Fermi
momenta $p_{+}¥$ and $p_{-}¥$.  The Green functions near ${\bf
p}=p_{\pm}¥$ and $\varepsilon=\mu_{\pm}¥$ have the form
\begin{equation}
G_{\pm}¥({\bf p}, \varepsilon)=\frac{a}{\varepsilon -
\mu_{\pm}¥-v_{F}¥(p-p_{\pm}¥) +ib(p-p_{\pm}¥)|p-p_{\pm}¥|}.
\label{e1}
\end{equation}

We shall assume a weak polarization
$v_{F}¥(p_{+}¥-p_{-}¥)\ll\varepsilon_{F}¥$ and also that both the Fermi
distributions are characterized by the same Landau Fermi liquid
parameters.  Unlike \cite{20} we introduce here the general form of
imaginary part of self-energy \cite{21} which is quadratic function of
the difference $(p-p_{\pm}¥)$ and changes its sign at $p=p_{\pm}¥$
correspondingly.  The assumption of small polarization in particular
means that $G_{+}¥$ is given by expression (\ref{e1}) not only near
its own Fermi surface $|{\bf p}|=p_{+}¥$ and $\varepsilon=\mu_{+}¥$ but
in whole intervals $p_{-}¥<p<p_{+}¥$ and
$\mu_{-}¥<\varepsilon<\mu_{+}¥$ and also near the "alien" Fermi
surface $|{\bf p}|=p_{-}¥$ and $\varepsilon=\mu_{-}¥$.  The same is
true for $G_{-}¥$.

In general the polarization is
nonequilibrium, hence $\mu_{+}¥-\mu_{-}¥=\Omega-\omega_{L}¥$, where
$\omega_{L}¥=\gamma H_{0}¥$ is the Larmor frequency corresponding to
the external field $H_{0}¥$ and $\Omega$ is the Larmor frequency
corresponding to the effective field which would produce the existing
polarization , $v_{F}¥(p_{+}¥-p_{-}¥)= \Omega/(1+F_{0}¥^{a}¥)$.
\cite{22}

Following Landau let us write equation for the vortex function for the
scattering of two particles with opposite spin direction and a small
transfer of 4-momentum $K=({\bf k},\omega)$
\begin{equation}
\Gamma(P_{1}¥,P_{2}¥,K)=\Gamma_{1}¥(P_{1}¥,P_{2}¥)-\frac{i}{(2\pi)^{4}¥}
\int \Gamma_{1}¥(P_{1}¥,Q)G_{+}¥(Q)G_{-}¥(Q+K)\Gamma(Q,P_{2}¥,K)d^{4}¥Q
\label{e2}
\end{equation}
If $K$ is small and polarization is also small, the poles of two Green
functions are close each other.
Let us assume that
all other quantities in the integrand are slowly varying with respect
to $Q$: their energy and momentum scales of variation are larger than $\max
\{\Omega, \omega\}$ and $\max\{\Omega/v_{F}¥,k\}$ correspondingly.
Then one can perform the integration in (\ref{e2}) at fixed values of 
$Q=p_{0}¥=(p_{+}¥+p_{-}¥)/2, ~\mu=(\mu_{+}¥+\mu_{-}¥)/2$ in the arguments
of $\Gamma$ and $\Gamma_{1}¥$ functions.  Another words, one can
substitute in (\ref{e2})
\begin{eqnarray}
&G_{+}¥&(Q)G_{-}¥(Q+K)=G_{+}¥({\bf q},\varepsilon)
G_{-}¥({\bf q}+{\bf k},\varepsilon+\omega) \nonumber \\
&=&\frac{2\pi i a^{2}¥}{v_{F}¥}\delta(\varepsilon-\mu)\delta(|{\bf q}|-p_{0}¥)
\frac{\frac{\Omega}{1+F_{0}¥^{a}¥}+{\bf k}{\bf v}_{F}¥}{\omega-\omega_{L}¥
+\frac{\Omega
F_{0}¥^{a}¥}{1+F_{0}¥^{a}¥}+\frac{ib\Omega^{2}¥}
{v_{F}¥^{2}¥(1+F_{0}¥^{a}¥)^{2}¥}- {\bf k}{\bf v}_{F}¥+ \frac{ib{\bf
k}{\bf v}_{F}¥\Omega}{v_{F}¥^{2}¥(1+F_{0}¥^{a}¥)}}+
\Phi_{\mbox{reg}}¥.
\label{e3}
\end{eqnarray}

For eliminating $\Gamma_{1}¥$ from (\ref{e2}) we shall rewrite this
equation in the operator form
\begin{equation}
\Gamma=\Gamma_{1}¥-i\Gamma_{1}¥(i\Phi+\Phi_{\mbox{reg}}¥)\Gamma,
\label{e4}    
\end{equation}
where product is interpreted as integral, and $i\Phi$ denotes the
first term from rhs eq.  (\ref{e3}).  In equation (\ref{e4}), we
transpose the term involving $\Phi_{\mbox{reg}}¥$ to the left-hand
side, and then apply the operator
$(1+i\Gamma_{1}¥\Phi_{\mbox{reg}}¥)^{-1}¥$, obtaining
\begin{equation}
\Gamma=\Gamma^{\omega}¥+\Gamma^{\omega}¥\Phi\Gamma,
\label{e5}
\end{equation}
where
\begin{equation}
\Gamma^{\omega}¥
=(1+i\Gamma_{1}¥\Phi_{\mbox{reg}}¥)^{-1}¥\Gamma_{1}¥.
\label{e6}
\end{equation}
As it is known \cite{14}, $\Gamma^{\omega}¥(\Omega=0)$
is directly related to the function
determining the Fermi liquid interaction,
\begin{equation}
\Gamma^{\omega}¥(\Omega=0)=\Gamma((|{\bf k}|/\omega)\to 0,\Omega=0)=
\frac{F_{{\bf n}{\bf n'}}}{a^{2}¥N_{0}¥}.
\label{e7}
\end{equation}
At finite $\Omega$ the $\Gamma^{\omega}¥$ function 
can be expanded over the polarization as
\begin{equation}
{a^{2}¥N_{0}¥}\Gamma^{\omega}¥={F_{{\bf n}{\bf n'}}} +
\frac{ib\Omega}{ v_{F}¥^{2}¥(1+F_{0}¥^{a}¥)}C_{{\bf n}{\bf n'}}
+O(\Omega^{2}¥).
\label{e8}
\end{equation}
From eqns (\ref{e5}) and (\ref{e8}), we come, according to well known
procedure \cite{14}, to kinetic equation
\begin{eqnarray}
&\left( \omega -\omega_{L}¥+\frac{\Omega
F_{0}¥^{a}¥}{1+F_{0}¥^{a}¥}+\frac{ib\Omega^{2}¥}
{v_{F}¥^{2}¥(1+F_{0}¥^{a}¥)^{2}¥} - {\bf k}{\bf n}v_{F}¥ + \frac{ib{\bf
k}{\bf n} v_{F}¥\Omega}{v_{F}¥^{2}¥(1+F_{0}¥^{a}¥)}\right )
\nu({\bf n})  \nonumber \\
&= \left (\frac{\Omega}{1+F_{0}¥^{a}¥}+{\bf k}{\bf
n}v_{F}¥\right) {\displaystyle \int} \frac{d{\bf n'}¥}{4\pi}
\left(F_{{\bf n}{\bf n'}}+
\frac{ib\Omega}{v_{F}¥^{2}¥(1+F_{0}¥^{a}¥)}C_{{\bf n}{\bf n'}}
\right)\nu({\bf n'}).
\label{e9}
\end{eqnarray}
We limit ourself by the first two harmonics in the Landau interaction
function $F_{{\bf n}{\bf n'}}=F_{0}¥^{a}¥+({\bf n}{\bf
n'})F_{1}¥^{a}¥$ and $C_{{\bf n}{\bf n'}}=C_{0}¥+({\bf n}{\bf
n'})C_{1}¥$.  To obtain the spectrum of the spin waves (see below)
obeying the Larmor theorem: the system of the spins in a homogeneous
magnetic field executes the precessional motion with the Larmor
frequency $\omega_{L}¥=\gamma H_{0}¥$, the coefficient $C_{0}¥$
has to be chosen equal to unity.

Introducing the expansion of the distribution
function $\nu({\bf n})¥$ over spherical harmonics of direction ${\bf
n}={\bf v}_{F}¥/v_{F}¥$ one can find from (\ref{e9}) that the ratio of
amplitudes of the successive harmonics with $l\ge 1$ is of the order
of $kv_{F}¥/\Omega$.  Hence if it is assumed this ratio as a small
parameter one can work with distribution function taken in the form
\cite{23} $\nu({\bf n})=\nu_{0}¥+({\bf n}\hat{\bf k})\nu_{1}¥$.  The
functions $\nu_{0}¥$ and $\nu_{1}¥$ obey the following system of
linear equations:
\begin{equation}
( \omega-\omega_{L}¥ )\nu_{0}¥-\frac{kv_{F}¥}{3}
\left (1+\frac{F_{1}¥^{a}¥}{3}
-\frac{ib(1-C_{1}¥/3)\Omega}{v_{F}¥^{2}¥(1+F_{0}¥^{a}¥)} \right ) \nu_{1}¥=0,
\label{e10}
\end{equation}
\begin{equation}
-kv_{F}¥(1+F_{0}¥^{a}¥)\nu_{0}¥+ \left
(\omega-\omega_{L}¥+\frac{\Omega
(F_{0}¥^{a}¥-\frac{F_{1}¥^{a}¥}{3})}{1+F_{0}¥^{a}¥}+
\frac{ib(1-C_{1}¥/3)\Omega^{2}¥}{v_{F}¥^{2}¥(1+F_{0}¥^{a}¥)^{2}¥}\right
)\nu_{1}¥=0.
\label{e11}
\end{equation}
The equality to zero of the determinant of this system gives the spin
waves dispersion law.  At long enough wave lengths when the dispersive
part of $\omega(k)$ dependence is much less than $\omega_{L}¥$ we have
\begin{equation}
\omega=\omega_{L}¥+ (D^{\prime\prime}¥-iD^{\prime}¥)k^{2}¥,
\label{e12}
\end{equation}
where 
\begin{equation} 
D^{\prime\prime}¥=\frac{v_{F}¥^{2}¥(1+F_{0}¥^{a}¥)(1+F_{1}¥^{a}¥/3)}
{3 \kappa\gamma H}
\label{e13}
\end{equation}
is the reactive part of diffusion coefficient,
\begin{equation}     
D^{\prime}¥=\frac{b(1-C_{1}¥/3) (1+F_{0}¥^{a}¥)^{2}¥}{3\kappa^{2}¥}
\label{e14}
\end{equation}
is dissipative part of diffusion coefficient,
$\kappa=F_{0}¥^{a}¥-F_{1}¥^{a}¥/3$ and
$H=\Omega/\gamma(1+F_{0}¥^{a}¥)$ is  effective "internal" field
corresponding to effective "external" field $\Omega/\gamma$ producing
the existing polarization.  We derived eqns (\ref {e13}) and (\ref {e14})
in the assumption of $\kappa \ne 0$.

The expressions for $D^{\prime}¥$ and $D^{\prime\prime}¥$ have been
obtained first by the same method by A.Meyerovich and Musaelyan
\cite{6}.  The former is literally coincides with found in this paper,
the latter has the same parametric dependence but depends in different
way from Fermi liquid parameters.  The reason for this is not clear at
the moment.
These expressions reproduce the corresponding diffusion constants have been
obtained from phenomenological Landau-Silin kinetic equation with
two-particle collision integral \cite{10} at arbitrary relation between
polarization and temperature if we put in the latters $T=0$.  In
particular $D^{\prime}¥$ proves to be polarization independent whereas
$D^{\prime\prime}¥$ is inversely proportional to polarization.

Thus, the general microscopic derivation confirms the statement about
the existance of zero-temperature spin waves attenuation in polarized
Fermi liquid.  The value of the dissipative part of spin diffusion
$D^{\prime}¥$ is determined by the amplitude "b" of the imaginary part of
self-energy.  Hence it originates of collisions between
quasiparticles.

\section{Fermi liquid approach to the "ferromagnetic" state}

There are known several attempts to consider an isotropic itinerant
ferromagnetic state as some peculiar type of Fermi liquid.
This subject
has been discussed first phenomenologically by A.A.Abrikosov and
I.E.Dzyaloshinskii \cite{11} and after microsopically by
P.S.Kondratenko \cite{12}.  They did not include in the theory a
finite scattering rate between quasiparticles and as result they have
obtained the dissipationless transvese spin wave dispersion law as it
seemed to be in isotropic ferromagnet.  The derivation \cite{11} has
been critisized by C.Herring \cite{24} who pointed out on the
existance of the finite scattering rate and inapplicability of naive
Fermi-liquid approach to itinerant ferromagnet (see also discussion in
\cite{10}).  Later I.E.Dzyaloshinskii and P.S.Kondratenko \cite{13}
have rederived the spin-wave dispersion law in ferromagnets.  Making
use as the starting point the Landau equation for the vertex function
for the scattering of two particles with opposite spin direction and a
small transfer of 4-momentum they have redefined the product of two
Green functions $G_{+}¥G_{-}¥$ in such a manner that its resonant part
was taken equal to zero at $\omega=0$.  This trick gives a possibility
to use the $1/k^{2}¥$ divergency of transverse static susceptibility
which is an inherent property of degenerate systems and occurs both in
an isotropic ferromagnet and in spin polarized paramagnetic
Fermi-liquid.  The latter of course is true in absense of interactions
violating of total magnetization conservation.  As in previous papers
\cite{11,12} the authors of \cite{13} did not introduce a scattering
rate in the momentum space between the Fermi surfaces for the
particles with opposite spins.

Let us see now what kind modifications appeare if we
reproduce  the derivation proposed in \cite{13} with the Green
functions (\ref{e1}) taking into account the finite quasiparticle
scattering rate in whole interval $p_{-}¥<p<p_{+}¥$. 
We discuss first an isotropic ferromagnet at equilibrium
$\mu_{+}¥=\mu_{-}¥$ in the absence of external field. Following
\cite{13} we write:
\begin{eqnarray}
&G_{+}¥&(Q)G_{-}¥(Q+K)=G_{+}¥({\bf q},\varepsilon)
G_{-}¥({\bf q}+{\bf k},\varepsilon+\omega) \nonumber \\
&=&\frac{2\pi i a^{2}¥}{v_{F}¥}\delta(\varepsilon-\mu)\delta(|{\bf q}|-p_{0}¥)
\frac{\omega} {\omega -v_{F}¥\Delta+ ib\Delta ^{2}¥ - {\bf k}{\bf
v}_{F}¥+ \frac{ib{\bf k}{\bf v}_{F}¥\Delta}{v_{F}¥}}+ \tilde
\Phi_{\mbox{reg}}¥,
\label{e15}
\end{eqnarray}
where $\Delta=p_{+}¥-p_{-}¥$.
Now the eqn (\ref{e2}) is written as
\begin{equation}
\Gamma=\Gamma_{1}¥-i\Gamma_{1}¥(i\tilde \Phi+\tilde \Phi_{\mbox{reg}}¥)\Gamma,
\label{e16}
\end{equation}
where $i\tilde \Phi$ denotes the first term from rhs eq.  (\ref{e15}). 
The equivalent form of this equation is
\begin{equation}
\Gamma=\Gamma^{\bf k}¥+\Gamma^{\bf k}¥\tilde\Phi\Gamma,
\label{e17}
\end{equation}
where
\begin{equation}
\Gamma^{\bf k}¥=\Gamma\left (\frac{\omega}{|{\bf k}|}\to 0\right )
=(1+i\Gamma_{1}¥\tilde\Phi_{\mbox{reg}}¥)^{-1}¥\Gamma_{1}¥.
\label{e18}
\end{equation}
The isotropic part of $\Gamma^{\bf k}¥$ is proportional to the
transversal susceptibility.  Hence it has the singular form \cite{13}
\begin{equation}     
\Gamma^{\bf k}¥\propto~
-\frac{1}
{N_{0}¥ (ck)^{2}¥ }.
\label{e19}
\end{equation}
Here, $c$ is constant with the dimensions of length.  It is quite
natural to take 
\begin{equation}  
    c\sim \frac{1}{\Delta}
\label{e20}
\end{equation}
such that the divergency (\ref{e19})
disappears in nonpolarized liquid when $\Delta~\to~0$.  The authors of \cite{13}
have loss this property by taking $c\sim p_{0}¥^{-1}¥$.

Substitution of eqn (\ref{e19}) to eqn (\ref{e17}) gives the
transversal spin waves dispersion law
\begin{equation}     
\omega=v_{F}¥\Delta(ck)^{2}¥( 1-\frac{ib\Delta}{v_{F}¥}),
\label{e21}
\end{equation}
which proves to be attenuating similar to the polarized Fermi-liquid.
One can take into consideration a static external field, by working in
the rotating with Larmor frequency coordinate frame that is equivalent
to the substitution $\omega \to \omega-\omega_{L}¥$ (see also \cite{13}).  As
result we obtain the law of dispersion
\begin{equation}     
\omega=\omega_{L}¥+v_{F}¥\Delta(ck)^{2}¥( 1-\frac{ib\Delta}{v_{F}¥}),
\label{e22}
\end{equation}
which obviously coincides with (\ref{e12}) after taking into account the
relation (\ref{e20}).

The attenuating dispersion of transversal spin waves is not surprising becouse
in both cases we deal in fact with spin polarized Fermi liquid but not
with isotropic itinerant ferromagnet where the zero temperature
atenuation is prohibited by Goldstone theorem.  The Fermi liquid
theory leading to the existance of such that attenuation is not
correct starting point for the construction of a theory of isotropic
itinerant ferromagnetism.

\section{Conclusion}

The inverse proportionality to polarization of reactive part diffusion
coefficient is typical for the polarized Fermi liquid.  It appears in
all the derivations of spin waves dispersion law including Fomin's
\cite{17}, Prange's \cite{18} and in Stoner-Hubbard model for
itinerant ferromagnetism (see for instance the book \cite{1}, or the
paper \cite{25} where the same results are obtained by more fashinable
now method of functional integration over fermi fields).  It has
nothing to do with dispersion law for a ferromagnet which must be
proportional to the magnetization as it follows from Landau-Lifshits
equation (\ref{ed}) taking into account the domain wall rigidity.  The latter
is the inherent property of ferromagnet and absent in the paramagnetic
polarized Fermi liquid.  The domain wall rigidity in itinerant
ferromagnet is formed becouse space time variations of momentum
dependent off-diagonal, or spin part of quasiparticle distribution
function are blocked up by the inevitable alteration of the orbital
part of many particle electron wave function being accompanyed by huge
increase of interaction energy.  It is not taken into account in Fermi
liquid theory.  From this point the famous Stoner-Hubburd model of
ferromagnetism overlooks the most important property of a
ferromagnet\ldots.  So, in our opinion the Fermi liquid theory is
applicable to spin-polarized Fermi liquid but not to the itinerant
ferromagnets.

In conclusion, we note that making use the quantum-field theoretical
approach, one can derive the dispersion law for the transverse spin
waves in a weakly polarized Fermi liquid at $T=0$.  Along with the
dissipationless part inversely proportional to the polarization it
contains also the finite zero-temperature damping.  The polarization
dependence both dissipative and reactive part of diffusion constants
corresponds to dependences found earlier by means of kinetic equation
with two-particle collision integral.  The same dispersion law is
derived by means of another approach where the divergency of the static
transverse susceptibility is taken into consideration.  These results
obtained for the system of fermions with Fermi liquid type ground state
are quite natural for spin polarized paramagnetic Fermi liquid.  
On the other hand in the isotropic itinerant ferromagnet one can expect
the dissipationless spin wave spectrum with reactive diffusion
constant proportional to magnetization.
This demonstrates the troubles of the Fermi liquid formulation of
theory of itinerant ferromagnetic systems which has to operate
with an ordered type of ground state.

\section{Acknowledgements}

This paper appeared as the answer to the question addressed me by
G.Jackeli and I express him my gratitude.
I am grateful to A. Meyerovich who kindly pointed me out on my
invalid pretension in the first version of this paper.  I indebted to
I.Fomin and W.Mullin for numerous discussions and E.Kats for the
interest to the subject.


\end{document}